\documentclass[12pt, oneside]{article}   	
\usepackage{geometry} 
\usepackage{amsmath} 								
\usepackage{graphicx}	
\usepackage{amssymb}
\usepackage{cite}
\usepackage{color,xspace}
\usepackage{extarrows}

\numberwithin{equation}{section}
\DeclareSymbolFont{extraup}{U}{zavm}{m}{n}
\DeclareMathSymbol{\vardiamond}{\mathalpha}{extraup}{87}
\usepackage{float}
\restylefloat{table}
\allowdisplaybreaks

\def\twomat[#1,#2][#3,#4]{\left( \begin{array}{cc} #1 & #2 \\ #3 & #4 \end{array} \right)}
\def\thv[#1,#2,#3]{\left( \begin{array}{c} #1 \\ #2 \\ #3 \end{array} \right)}
\def\twv[#1,#2]{\left( \begin{array}{c} #1 \\ #2 \end{array} \right)}
\def\lagrange{\mathcal{L}}
\def\nn{\nonumber}

\title{R-symmetry for Higgs alignment without decoupling}

\date{}

\begin{document}

\begin{flushright}
\end{flushright}
\begin{center}

\vspace{1cm}
{\LARGE{\bf R-symmetry for Higgs alignment without decoupling}}

\vspace{1cm}

\large{\bf Karim Benakli$^\spadesuit$ \let\thefootnote\relax\footnote{$^\spadesuit$kbenakli@lpthe.jussieu.fr}
Yifan Chen$^{\vardiamond}$ \let\thefootnote\relax\footnote{$^\vardiamond$yifan.chen@lpthe.jussieu.fr}
 Ga\"etan~Lafforgue-Marmet$^\clubsuit$ \footnote{$^\clubsuit$glm@lpthe.jussieu.fr}
 \\[5mm]}

%{\small
%\emph{1-- Sorbonne Universit\'es, UPMC Univ Paris 06, UMR 7589, LPTHE, F-75005, Paris, France \\
%2-- CNRS, UMR 7589, LPTHE, F-75005, Paris, France }}
{ \sl Laboratoire de Physique Th\'eorique et Hautes Energies (LPTHE),\\ UMR 7589,
Sorbonne Universit\'e et CNRS, 4 place Jussieu, 75252 Paris Cedex 05, France.}

\end{center}
\vspace{0.7cm}

\abstract{It has been observed that an automatic alignment without decoupling is predicted at tree-level in a Two-Higgs Doublet Model (2HDM) with extended supersymmetry in the gauge and Higgs sectors. Moreover, it was found that radiative corrections preserve this alignment to a very good precision. We show that it is the non-abelian global $SU(2)_R$ R-symmetry that is at the origin of this alignment. This differs from previously considered Higgs family  symmetries as it is present only in the quartic part of the Higgs potential. It can not be imposed to the quadratic part  which has to be generated by $N=1$ supersymmetry breaking sectors. This absence of symmetry does not spoil alignment at the minimum of the potential. We show how the (small) misalignment induced by higher order corrections can be described as the appearance of non-singlet representations of $SU(2)_R$ in the quartic potential. }

\newpage
\setcounter{footnote}{0}

%------------------------------------------------------------------------------------------------------------							
\section{Introduction}
\label{introduction}

%------------------------------------------------------------------------------------------------------------

In contrast to fermions and vectors, there is only one known fundamental spin zero particle in Nature: the Standard Model (SM) Higgs boson. Additional fundamental scalars are ubiquitous in Early Universe cosmological and supersymmetric models. In the latter, matter fermions have scalar partners.  Also, additional Higgs scalars appear in their electroweak symmetry breaking sectors. Mixings of the observable Higgs with the new scalars are subject to strong constraints by present LHC experiments data. They imply that the observed Higgs, an eigenstate of the scalars mass matrix, is \emph{aligned} with the direction acquiring a non-zero vacuum expectation value (v.e.v).  Making all the additional scalars heavy enough, thus decoupling them, is a trivial way to achieve this. But a more interesting option is that alignment emerges as a consequence of specific patterns of the model. The benefits are that less constraints on masses allow to keep new scalars within the reach of future searches at the LHC. Such an  \emph{alignment without decoupling} was obtained in \cite{Antoniadis:2006uj} and further discussed in \cite{Ellis:2016gxa,Benakli:2018vqz}.

The effective low energy scalar potential of \cite{Antoniadis:2006uj}, studied in details in \cite{Belanger:2009wf}, corresponds to a peculiar case of a Two Higgs Doublet Model (2HDM)  \footnote{For an introduction to 2HDM see for example \cite{Gunion:1989we,Branco:2011iw,Djouadi:2005gj}}.  Symmetries in the 2HDM ( e.g. \cite{Davidson:2005cw,Ivanov:2005hg,Ferreira:2009wh}) can in particular cases imply alignment without decoupling \cite{Gunion:2002zf,Dev:2014yca,Lane:2018ycs}. Unfortunately, these symmetries have been quoted to lead to problematic phenomenological consequences, as massless quarks \cite{Ferreira:2010bm}. When not due to symmetries, alignment without decoupling remains viable. This situation was discussed for example in \cite{Bernon:2015qea,Bernon:2015wef,Carena:2013ooa,Carena:2015moc,Haber:2017erd} for the MSSM and NMSSM.  This remains not totally satisfactory as it is due to an ad-hoc specific choice of the model parameters. It is to be contrasted with \cite{Antoniadis:2006uj} where the alignment at tree-level is a prediction. It is the purpose of this work to uncover the symmetry at the origin of the automatic Higgs alignment in \cite{Antoniadis:2006uj}, a success that was shown to survive with a impressive precision when radiative corrections, up to two-loop, are taken into account \cite{Benakli:2018vqz}.

The crucial ingredient in  \cite{Antoniadis:2006uj} is the presence of $N=2$ extended supersymmetry. Early models have required that $N=2$ supersymmetry acts on the whole SM states and, as a consequence, suffered from the non-chiral nature of quarks and leptons \cite{Fayet:1975yi,delAguila:1984qs}.  Overcoming this issue by allowing both $N=2$ supersymmetry and chirality is possible in models inspired by superstring theory: orbifold fixed points and brane localizations enable to construct models where different parts preserve different amounts of supersymmetries. In \cite{Antoniadis:2006uj} the (non-chiral) gauge and Higgs states appear in a $N=2$ supersymmetry sector while the matter states, quarks and leptons, appear in an $N=1$ sector. An important feature of such constructions considered here \cite{Antoniadis:2005em,Antoniadis:2006eb,Antoniadis:2006uj,Allanach:2006fy,Belanger:2009wf} is that  gauginos have Dirac masses \cite{Fayet:1978qc,Polchinski:1982an,Hall:1990hq,Fox:2002bu,Benakli:2008pg}. Also these $N=2$ extended models have implication for Higgs boson physics as discussed in \cite{Belanger:2009wf,Benakli:2009mk,Amigo:2008rc,Benakli:2010gi,Choi:2010gc,Benakli:2011vb,Benakli:2011kz,Itoyama:2011zi,Benakli:2012cy,Benakli:2014cia,Martin:2015eca,Braathen:2016mmb,Unwin:2012fj,Chakraborty:2018izc,Csaki:2013fla}.

It is not totally satisfactory to explain the predicted alignment only by the presence of $N=2$ extended supersymmetry because this is realized only at a very high energy, the fundamental scale of the theory, while it should be possible to trace back the alignment to a symmetry manifest in the scalar potential of 2HDM. In this work, we will exhibit the relevant symmetry.

The paper is organized as follows. In Section 2, the main ingredients of the model are presented succinctly. This allows to define the notation used through this work. In section 3, we show that the potential can be written as a sum of two $SU(2)_R$ R-symmetry singlet representations. We  also show how this implies alignment. In section 4, we review how higher order corrections induce a small misalignment.  Section 5 presents our conclusions.

\newpage

%--------------------------------------------------------------
\section{A glimpse of the model }
%\label{2HDM_Limit}
%---------------------------------------------------------

%%%%%%%%%%%%%%%%%%%%%%%%

The $N=2$ structure of the gauge sector implies the presence of chiral superfields in the adjoint representations of SM gauge group. These are a singlet $\mathbf{S}$ and an $SU(2)$ triplet  $\mathbf{T}$. We define
\begin{eqnarray}
S &=& \frac{S_R + iS_I}{\sqrt{2}} \\
T &=& \frac{1}{2} 
\begin{pmatrix}
T_0 & \sqrt{2} T_+ \\
\sqrt{2}T_- & -T_0 
\end{pmatrix} \,, \qquad T_i = \frac{1}{\sqrt{2}}\, (T_{iR} + i T_{iI})  \quad {\rm  with}  \quad i=0,+, -
\end{eqnarray}

Relevant to this work is the fact that the adjoint fields contribute to the superpotential  by promoting the gauginos to Dirac fermions, but also by generating new Higgs interactions through:

\begin{eqnarray}
W = && \sqrt{2} \, \mathbf{m}_{1D}^\alpha \mathbf{W}_{1\alpha} \mathbf{S} + 2 \sqrt{2} \, \mathbf{m}_{2D}^\alpha \text{tr} \left( \mathbf{W}_{2\alpha} \mathbf{T}\right)  + \frac{M_S^2}{2} \mathbf{S}^2 + \frac{\kappa}{3} \mathbf{S}^3 + M_T \, \text{tr} (\mathbf{TT}) \,  \nonumber \\ 
 && + \mu \,  \mathbf{H_u} \cdot \mathbf{H_d}+ \lambda_S S \, \mathbf{H_u} \cdot \mathbf{H_d} + 2 \lambda_T \, \mathbf{H_d} \cdot \mathbf{T H_u} \,,
\end{eqnarray} 

\noindent where the Dirac masses can be read by taking $\mathbf{m}_{\alpha i D} := \theta_\alpha m_{iD} $ where $\theta_\alpha$ are the Grassmanian superspace coordinates. We also need to define the soft terms  appearing in the Higgs and adjoint scalar sectors,  chosen for simplicity to be real,
\begin{align}
\lagrange_{\rm soft} =& m_{H_u}^2 |H_u|^2 + m_{H_d}^2 |H_d|^2 + B{\mu} (H_u \cdot H_d + \text{h.c})  \nonumber \\ 
 & + m_S^2 |S|^2 + 2 m_T^2 \text{tr} (T^{\dagger} T) + \frac{1}{2} B_S \left(S^2 + h.c\right)+  B_T\left(\text{tr}(T T) + h.c.\right)  \label{soft} \\
& + A_S \left(S H_u \cdot H_d + h.c \right) + 2 A_T  \left( H_d \cdot T H_u + h.c \right) + \frac{A_\kappa}{3} \left( S^3 + h.c. \right) + A_{ST} \left(S \mathrm{tr} (TT) + h.c \right). \nn 
\end{align} 

\noindent Considering specific scenarios  for generation of the above soft-terms leads us to make some assumption about their relative sizes. In the case of gauge mediation, one is first tempted to consider secluded supersymmetry breaking and the mediators sectors  to have an $N=2$ structure \cite{Fayet:1975yi}. This would allow to preserve the underlying $R$-symmetry. Unfortunately, it leads to tachyonic directions for the adjoint scalars. To overcome this, we must restrict to particular $N=1$ breaking and mediating sectors \cite{Benakli:2008pg,Benakli:2016ybe,Csaki:2013fla} (see also \cite{Amigo:2008rc,Nelson:2015cea,Alves:2015kia,Alves:2015bba}). As a consequence, the $SU(2)_R$ structure is not preserved by the quadratic part of the scalar potential given by the soft breaking terms. Furthermore, we will consider that the remaining  $U(1)_R$ symmetry is broken by the presence of a non-vanishing $B\mu$ term (keeping zero the coefficients of supersymmetric terms as $M_S$, $M_T$,  and $M_O$). We also take $\kappa =0$ for simplicity. This avoids then the introduction of extra-doublets as in the MRSSM \cite{Amigo:2008rc} and allows us to consider an effective 2HDM. Note also that the trilinear terms in the last line of (\ref{soft}) will be neglected here, as they were shown to be generically small \cite{Benakli:2008pg,Benakli:2016ybe}.

The minimization of this potential was discussed in \cite{Belanger:2009wf}.  An effective 2HDM is obtained by taking the limit $m^2_S, m_T^2 \gg m_Z^2$ and keeping the Dirac masses $m_{iD}$ as well as $m_{H_u}^2, m_{H_d}^2, B\mu$ and $\mu$ small.  Note that $B\mu$ measures the size of $N=1$ $U(1)_R$ symmetry breaking. The effective theory just above the electroweak scale is a 2HDM with a set of light charginos and neutralinos. The minimization of the corresponding potential was discussed in \cite{Belanger:2009wf}. We restrict to the case of CP neutral vacuum, i.e. $H^0_{dI}=H^0_{uI} =0$ which implies also $S_I=T_I=0$. Dropping the obvious $0$ indices for neutral components, we define:
\begin{eqnarray}
M_Z^2 & =& \frac{g_Y^2 + g_2^2}{4} v^2 \, ,  \qquad  v \simeq 246 {\rm GeV}  \\
<H_{uR}>& =& {v s_\beta }, \qquad <H_{dR}>={v c_\beta},  \\
 <S_R>&=& v_s \, , \qquad <T_R>=v_t
\end{eqnarray}
where:
\begin{eqnarray}
c_\beta &\equiv& \cos \beta,\qquad   s_\beta \equiv \sin \beta, \qquad  t_\beta \equiv \tan\beta \, ,  \qquad   0 \leqslant \beta \leqslant \frac{\pi}{2} \nonumber \\
c_{2\beta} &\equiv& \cos 2\beta,\qquad   s_{2\beta} \equiv \sin 2\beta
\end{eqnarray}

Also, the leading-order squared-masses for the real part of the adjoint fields are given by \cite{Benakli:2011kz}:
\begin{align}
m_{SR}^2 =& m_S^2 + 4 m_{DY}^2 + B_S , \qquad m_{TR}^2 = m_T^2  + 4 m_{D2}^2 + B_T \, .
\label{STscalarmasses}
\end{align}

Note that as we have
\begin{eqnarray}
v_s &\simeq &\frac{v^2}{2 m_{SR}^2} \ \ { \left[\, \, \, \, \,  g_Y    m_{1D} c_{2\beta} + {\sqrt{2}} \mu    \lambda_S    \right]} \nonumber \\
v_t & \simeq &  \frac{v^2}{2 m_{TR}^2} \ \ \left[ - g_2    m_{2D}  c_{2\beta} -{\sqrt{2}} \mu    \lambda_T  \right], 
\end{eqnarray}
the bounds on the expectation value of the DG-triplet come from the electroweak precision data, i.e. the contribution to the $\rho$ parameter, remains acceptable in our scenario as we keep $m_{iD}$ and $\mu$ an order of magnitude smaller than $m_{SR}$ and $m_{TR}$: the light charginos and neutralinos are in the sub-TeV energy region accessible to future searches.

%--------------------------------------------------------------
\section{$R$-symmetric Higgs alignment}
%\label{2HDM_Limit}
%---------------------------------------------------------

The model has a $U(1)_R \times SU(2)_R$ global R-symmetry. The $U(1)_R$ is the R-symmetry present at the level of $N=1$ and plays no role in what follows. Using
\begin{align}
\Phi_2 = H_u, \qquad \Phi_1^i = -\epsilon_{ij} (H_d^j)^* \Leftrightarrow \twv[H_d^0,H_d^-] = \twv[\Phi_1^0,-(\Phi_1^+)^*] 
\end{align}
the two Higgs doublets can be assembled into one hypermultiplet $(\Phi_1, \Phi_2)^T$ in the fundamental representation of the $SU(2)_R$ R-symmetry.
The coupling of vector multiplet fields to hypermultiplet leads then to the presence of additional terms in the superpotential 
\begin{eqnarray}
W_{\text{Higgs}} \supset  \lambda_S \mathbf{S} \, \mathbf{\Phi_2}^\dagger \cdot \mathbf{\Phi_1} + 2 \lambda_T \, \mathbf{\Phi_1}^\dagger \cdot \mathbf{T} \mathbf{\Phi_2} \label{W_Higgs}
\end{eqnarray} 
and $N=2$ supersymmetry requires 
\begin{align}
\lambda_S= \frac{1}{\sqrt{2}} g_Y, \qquad \lambda_T =\frac{1}{\sqrt{2}} g_2
\label{LSTN2}
\end{align}
where $g_Y$ and $g_2$ stand for the hypercharge and $SU(2)$ gauge couplings, respectively.

The integration out of adjoint scalars leads to a potential for the Higgs fields that corresponds to a peculiar 2HDM. Generic 2HDM are usually parametrized as: 
\begin{eqnarray}
V_{EW} &=& V_{2\Phi} +V_{4\Phi} 
\label{decomp2HDM} 
\end{eqnarray}
where
\begin{eqnarray}
V_{2\Phi} &=& m_{11}^2 \Phi_1^\dagger \Phi_1 + m_{22}^2 \Phi_2^\dagger \Phi_2 - [m_{12}^2 \Phi_1^\dagger \Phi_2 + \text{h.c}] \nonumber \\
V_{4\Phi} &=& \frac{1}{2} \lambda_1 (\Phi_1^\dagger \Phi_1)^2 + \frac{1}{2} \lambda_2 (\Phi_2^\dagger \Phi_2)^2 \nonumber \\
& & +  \lambda_3(\Phi_1^\dagger \Phi_1) (\Phi_2^\dagger \Phi_2) + \lambda_4 (\Phi_1^\dagger \Phi_2)(\Phi_2^\dagger \Phi_1) \nonumber \\ 
& & + \left[ \frac{1}{2} \lambda_5 (\Phi_1^\dagger \Phi_2)^2 + [\lambda_6 (\Phi_1^\dagger \Phi_1) + \lambda_7 (\Phi_2^\dagger \Phi_2)] \Phi_1^\dagger \Phi_2 + \text{h.c} \right]\,,  
\label{reparametriz2HDM} \,
\end{eqnarray}
from which we can write down
\begin{eqnarray}
m_{11}^2 &=& m_{H_{d}}^2 + \mu^2, \qquad m_{22}^2 = m_{H_{u}}^2 + \mu^2, \qquad m_{12}^2 = B\mu . 
\label{2HDM_params}
\end{eqnarray}

The parameters $\lambda_i$ can be decomposed as their leading order tree-level values and  corrections due to threshold $\delta \lambda_i^{(tree)} $ and loops $\delta \lambda_i^{(rad)}$
\begin{align}
\lambda_i =&\lambda_i^{(0)} + \delta \lambda_i^{(tree)} + \delta \lambda_i^{(rad)}
\label{deltaLs}
\end{align}
whose values were computed in \cite{Belanger:2009wf,Benakli:2012cy}.

In our minimal model:
\begin{align}
\lambda_5 =& \lambda_6 = \lambda_7 =0.
\label{lambdaZeros}
\end{align}
while
\begin{align}
\lambda_1^{(0)}= \lambda_2^{(0)} =& \frac{1}{4} (g_2^2 + g_Y^2)  \nn \\
\lambda_3^{(0)} =&   \frac{1}{4}(g_2^2 - g_Y^2) + 2 \lambda_T^2  \qquad \xlongrightarrow{N=2}  \qquad  \frac{1}{4}(5 g_2^2 - g_Y^2)  \nn  \\
\lambda_4 ^{(0)}=& -\frac{1}{2}g_2^2 + \lambda_S^2 - \lambda_T^2  \qquad \xlongrightarrow{N=2} \qquad - g_2^2 + \frac{1}{2} g_Y^2\nn\\
\label{EQ:MDGSSMTree}\end{align}

At this leading order, the quartic part of the potential can then be put in the form:
\begin{eqnarray}
V_{4\Phi} &=&\frac{ 3 g_2^2}{8}  \left[  (\Phi_1^\dagger \Phi_1) + (\Phi_2^\dagger \Phi_2) \right]^2  
+   \frac{(- 2 g_2^2 +  g_Y^2)}{8}   \left[  \left( (\Phi_1^\dagger \Phi_1) - (\Phi_2^\dagger \Phi_2) \right)^2 +4 (\Phi_1^\dagger \Phi_2)(\Phi_2^\dagger \Phi_1)   \right]  \nonumber \\
 &=& \, \, \lambda_{|0_1,0>}   |0_1,0\rangle  \quad +  \quad  \lambda_{|0_2,0>}   |0_2,0\rangle
\label{reparam2HDM-SU2R} 
\end{eqnarray}
where we recognize between brackets the two  invariant combinations under $SU(2)_R$. These quartic terms, thus product of four fields, are the combination of the two $SU(2)_R$  doublets giving singlet irreducible representations. In the last equality, they are written in the standard spin representation notation  $|l,m>$ with  $l$  the spin and $m$ its projection along the $z$ axis. They were classified in \cite{Ivanov:2005hg}. Here \footnote{We found a minus sign for the $|0_2,0\rangle$  but compensated by a minus sign in $\lambda_{|0_2,0>} $ compared to \cite{Ivanov:2005hg}.}:
\begin{eqnarray}
\begin{array}{lll}
|0_1,0\rangle&=& \frac{1}{2}\left[(\Phi^\dagger_1\Phi_1) + (\Phi^\dagger_2 \Phi_2)\right]^2 \, ,
\end{array}
\label{1-1ofSU(2)}
\end{eqnarray}
\begin{eqnarray}
\begin{array}{lll}
|0_2,0\rangle&=& - \frac{1}{\sqrt{12}}\left[\left((\Phi^\dagger_1\Phi_1) - (\Phi^\dagger_2 \Phi_2)\right)^2 \, 
+ 4(\Phi^\dagger_2\Phi_1)(\Phi^\dagger_1\Phi_2)\right]
\end{array}
\label{1-2ofSU(2)}
\end{eqnarray}
while 
\begin{eqnarray}
\lambda_{|0_1,0>} &=&\frac {\lambda_1 + \lambda_2 + 2\lambda_3}{ 4} =  \frac{ 3 g_2^2}{4} 
\label{lambdaofSU(2)1}
\end{eqnarray}
and
%\qquad  {\rm  and}  \quad
\begin{eqnarray}
\lambda_{|0_2,0>} = -\frac  {\lambda_1 + \lambda_2 - 2\lambda_3 +4\lambda_4 }{ 4\sqrt{3}} = - \frac{\sqrt{3}(- 2 g_2^2 +  g_Y^2)}{4}\, .
\label{lambdaofSU(2)2}
\end{eqnarray}
Therefore, the $SU(2)_R$ R-symmetry  acts here as an $SU(2)$ Higgs family symmetry \cite{Davidson:2005cw,Ivanov:2005hg}. The potential contains only terms that are invariant (singlet) under $SU(2)_R$.

For the case of CP conserving Lagrangian under consideration, there are two CP even scalars with squared-mass matrix
\begin{eqnarray}
\mathcal{M}^2_h = \begin{pmatrix}
Z_1 v^2 & Z_6 v^2 \\
Z_6 v^2 & m_A^2 + Z_5 v^2 \end{pmatrix} \, . 
\label{2HDM_mass_matrix}
\end{eqnarray}
Using the notation $\lambda_{345} \equiv \lambda_3 + \lambda_4 + \lambda_5$, we have 
\begin{align}
Z_1 =& \lambda_1c_\beta^4 + \lambda_2 s_\beta^4 + \frac{1}{2} \lambda_{345} s_{2\beta}^2, \qquad \xlongrightarrow{N=2} \qquad \frac{1}{4} (g_2^2 + g_Y^2)  \nn\\
 Z_5 =& \frac{1}{4} s_{2\beta}^2 \left[ \lambda_1 + \lambda_2 - 2\lambda_{345}\right] + \lambda_5 \qquad \xlongrightarrow{N=2}  \qquad 0 \nn\\
Z_6 =& -\frac{1}{2} s_{2\beta} \left[\lambda_1 c_\beta^2 - \lambda_2 s_\beta^2 - \lambda_{345} c_{2\beta} \right] \qquad \xlongrightarrow{N=2} \qquad  0.
\label{2HDMZ}
\end{align} 
while  the pseudoscalar mass $m_A$ is given by
\begin{align}
m_A^2 =& - \frac{m_{12}^2}{s_\beta c_\beta} - \lambda_5 v^2 \qquad \xlongrightarrow{N=2} \qquad  - \frac{m_{12}^2}{s_\beta c_\beta}
\end{align}
Denoting by $M_Z$ and $M_W$ the  Z and W boson masses respectively,  we find that the $SU(2)_R$ symmetric potential leads to a diagonal  squared-mass matrix with eigenvalues:
\begin{align}
m_{h}^2 &= \frac{1}{4} (g_2^2 + g_Y^2) v^2= M_Z^2 \nn\\
 m_{H}^2 &= m_A^2 
\label{ZH}
\end{align} 
and a charged Higgs of mass
\begin{align}
m_{H^+}^2 =& \frac{1}{2} ( \lambda_5 - \lambda_4) v^2 + m_{A}^2 \qquad \xlongrightarrow{N=2} \qquad    \frac{1}{2}  (g_2^2 - \frac{1}{2} g_Y^2) v^2 + m_{A}^2=3M_W^2-M_Z^2+m_A^2.
\end{align}

It is a well known fact (e.g; \cite{Gunion:2002zf}) that $Z_6$ can be expressed, as we have written above as functions of only the quartic potential parameters, or as function of mass parameters. The different expressions are related by minimization conditions. We would like now to explain what this implies in our specific model for the mass parameters.

Let us turn now to the quadratic part of the potential. It can be written as:
\begin{eqnarray}
V_{2\Phi} &=& \, \,   \frac{ m_{11}^2  + m_{22}^2}{\sqrt{2}}\times  \frac{1}{\sqrt{2}} \left[  (\Phi_1^\dagger \Phi_1) + (\Phi_2^\dagger \Phi_2)  \right]
 \nonumber \\
& & +  \frac{ m_{11}^2  - m_{22}^2}{\sqrt{2}}\times  \frac{1}{\sqrt{2}} \left[  (\Phi_1^\dagger \Phi_1) - (\Phi_2^\dagger \Phi_2)  \right] \nonumber \\
& & - [m_{12}^2 \Phi_1^\dagger \Phi_2 + \text{h.c}] 
\label{reparam2HDM} 
\end{eqnarray}

Only the first line corresponds to the $SU(2)_R$ invariant term. Imposing a Higgs family symmetry would have required that both coefficients of the two non-invariant operators to vanish, therefore $m_{11}^2  = m_{22}^2$ and $m_{12}=0$. This is problematic in our model. First, it implies $m_{A}^2=0$. 
There is also another reason why we do not want to impose invariance of the quadratic potential under $SU(2)_R$: these terms are generated by supersymmetry breaking effects. One possible scenario is gauge mediation. An $N=2$ supersymmetry breaking and messengers sectors has been shown to lead to tachyonic masses for the adjoint scalars $S$ and $T$ \cite{Benakli:2008pg}. Avoiding these instability requires peculiar structure of these sectors that is not compatible with the $R$-symmetry.

However, we will obtain some conditions on the coefficients of the quadratic terms from the minimization of the potential. The corresponding equations take the form ( e.g. \cite{Haber:1993an}):
\begin{eqnarray}
0 &=& m_{11}^2 - t_{\beta} m_{12}^2  + \frac{1}{2} v^2 c_\beta^2 (\lambda_1 +  \lambda_6 t_\beta + \lambda_{345} t_\beta^2 + \lambda_7 t_\beta^2)     \\
0 &=& m_{22}^2 - \frac{1}{t_{\beta}} m_{12}^2  + \frac{1}{2} v^2 s_\beta^2 (\lambda_2 +  \lambda_7 \frac{1}{t_\beta} + \lambda_{345} \frac{1}{t_\beta^2} + \lambda_6 \frac{1}{t_\beta^2})  
\label{EqtsMotion1} 
\end{eqnarray} 

Plugging  the values of $\lambda_i^{(0)}$ in (\ref{EqtsMotion1}) leads to the equations:

\begin{eqnarray}
0 &=& m_{11}^2 - t_{\beta} m_{12}^2  + \frac{1}{8} (g_2^2 + g_Y^2) v^2     \\
0 &=& m_{22}^2 - \frac{1}{t_{\beta}} m_{12}^2  + \frac{1}{8} (g_2^2 + g_Y^2) v^2  
\label{EqtsMotion2} 
\end{eqnarray} 
Subtraction of one of the equations from the other one leads (for $s_{2\beta} \neq 0$) to
\begin{eqnarray}
0 &=& \frac{1}{2} (m_{11}^2 - m_{22}^2) s_{2\beta} + m_{12}^2  c_{2\beta} \equiv Z_6 v^2
\label{EqtsMotion3} 
\end{eqnarray} 
We see that the the values of the quartic couplings imply that the $SU(2)_R$ violating terms are such that their combination contributing to $Z_6$ vanishes. Therefore, we have shown that the constraint of $SU(2)_R$ invariance of the quartic part of the potential is sufficient to insure an automatic alignment without decoupling. On the other hand, we also found that given $m_{11}^2$, $m_{22}^2$ and $ m_{12}^2$ the potential minimization equation fixes  $\beta$  such that an alignment is obtained. This is different from the \cite{Gunion:2002zf,Dev:2014yca} where  $\beta$ remains arbitrary.

These results also allow us to understand why the simple identification of couplings in the Higgs couplings with their $N=2$ expected values, as it was attempted for the MRSSM in \cite{Benakli:2018vqz}, fails to achieve alignment. There, the integration out of the additional doublets breaks the $SU(2)_R$ R-symmetry explicitly at tree level.

%--------------------------------------------------------------
\section{$R$-symmetry breaking and misalignment}
%\label{2HDM_Limit}
%---------------------------------------------------------

%%%%%%%%%%%%%%%%%%%%%%%%

We have shown how the alignment is enforced by the $SU(2)_R$ symmetry of the quartic potential. However, it is clear that the leading order values $\lambda_i^{(0)}$ received corrections from many sources that do not respect the $SU(2)_R$ symmetry. It was numerically checked, computing two-loop quantum corrections,  that the alignment remains amazingly true to a very high precision when taking into account the full corrections  \cite{Benakli:2018vqz}. We would like to investigate how these sub-leading terms affect not only the numerical value but the $SU(2)_R$ group theory structure of the scalar potential. The quartic scalar potential can then be written as:
\begin{eqnarray}
V_{4\Phi} &=& \, \, \sum_{i,j} \lambda_{|i,j>} \times   |i,j\rangle
\label{reparam2HDM-2} 
\end{eqnarray}
where $|i,j\rangle$ are irreducible representations of $SU(2)_R$.  

The alignment is measured by computing the off-diagonal squared-mass matrix element
\begin{align}
Z_6 = -\frac{1}{2} s_{2\beta} \left[\lambda_1 c_\beta^2 - \lambda_2 s_\beta^2 - \lambda_{345} c_{2\beta} \right] 
\label{Z6-tree}
\end{align} 
Interestingly, this can be recasted  as
\begin{align}
Z_6 = \frac{1}{2} s_{2\beta} \left[  \sqrt{2} \lambda_{|1,0>} -  \sqrt{6} \lambda_{|2,0>} c_{2\beta}  + (\lambda_{|2,-2>} + \lambda_{|2,+2>}) c_{2\beta}. \right] 
\label{Z6-tree-SU(2)}
\end{align} 
with, using the notation of \cite{Ivanov:2005hg}:
\begin{eqnarray}
\begin{array}{lll}
|1,0\rangle&=& \frac{1}{\sqrt{2}}\left[(\Phi^\dagger_2\Phi_2) - (\Phi^\dagger_1 \Phi_1)\right]
\left[(\Phi^\dagger_1\Phi_1) + (\Phi^\dagger_2 \Phi_2)\right] \\[1.5mm]
|2,0\rangle&=& \frac{1}{\sqrt{6}}\left[(\Phi^\dagger_1\Phi_1)^2 + (\Phi^\dagger_2 \Phi_2)^2 
 - 2(\Phi^\dagger_1\Phi_1)(\Phi^\dagger_2\Phi_2) -  2(\Phi^\dagger_1\Phi_2)(\Phi^\dagger_2\Phi_1)\right] \\[1.5mm]
|2,+2\rangle&=& (\Phi^\dagger_2\Phi_1)(\Phi^\dagger_2 \Phi_1) \\ [1.5mm]
|2,-2\rangle&=& (\Phi^\dagger_1\Phi_2)(\Phi^\dagger_1 \Phi_2) 
\end{array}
\label{5ofSU(2)}
\end{eqnarray}
The coefficients that play a role in the misalignment (\ref{Z6-tree-SU(2)}) are given as function of $\lambda_i$ by:
\begin{eqnarray}
\lambda_{|1,0>} &=&  \frac {\lambda_2 - \lambda_1 }{ 2\sqrt{2}}\, \qquad \xlongrightarrow{{\rm leading\,  order}}  \qquad  0 \nn \\ [1.5mm]
\lambda_{|2,0>}  &= &\frac {\lambda_1 + \lambda_2 - 2\lambda_3 - 2\lambda_4 }{ \sqrt{24}}\,\qquad \xlongrightarrow{{\rm leading\,  order}}  \quad -  \frac{1}{\sqrt{24}}  \, \left[ (2 \lambda_S^2- g_Y^2  )  + ( 2 \lambda_T^2 -g_2^2 ) \right] \nn  \\ [1.5mm]
\lambda_{|2,+2>} &=& \frac{\lambda_5^*}{2}\, \quad \xlongrightarrow{{\rm leading\,  order}}  \quad  0, \qquad 
 \lambda_{|2,-2>} = \frac{\lambda_5}{2}\, \quad \xlongrightarrow{{\rm leading\,  order}}  \quad  0.
\label{lambdaofSU(2)}
\end{eqnarray}
We find that breaking of the $SU(2)_R$ invariance in the quartic potential is necessary in order to have misalignment. The conservation of the $U(1)$ subgroup of $SU(2)_R$ is not sufficient for alignment as we have contribution from $|i,0\rangle$ combinations.
Here, we have $\lambda_5 =0$ thus there is no contribution from $|2,\pm 2>$. The breaking of $SU(2)_R$ symmetry leads then to a contribution to the $Z_6$ parameter of order:
\begin{align}
\delta  Z_6^{(tree)} = \frac{1}{2} s_{2\beta} \left[  \sqrt{2} \delta \lambda_{|1,0>} -  \sqrt{6} \delta   \lambda_{|2,0>} c_{2\beta} \right] 
\label{mis-Z6-tree-SU(2)}
\end{align} 
where $\delta \lambda_{|i,0>}$ are corrections generated by higher order corrections to the tree-level $ \lambda_{|i,0>}^{(0)}$.

First, the parameters $\lambda_i$ receive tree-level correction from the threshold when the adjoint scalars are integrated out while the Higgs $\mu$-term and the Dirac masses $m_{1D}, m_{2D}$ for the hypercharge $U(1)$ and weak interaction $SU(2)$ are small but not zero:
\begin{align}
\delta \lambda_1^{(tree)} \simeq & -\frac{\left(g_Y m_{1D} - \sqrt{2} \lambda_S \mu\right)^2}{m_{SR}^2} - \frac{\left(g_2 m_{2D} + \sqrt{2} \lambda_T  \mu\right)^2}{m_{TR}^2}\nn \\
\delta  \lambda_2 ^{(tree)} \simeq &   -\frac{\left(g_Y m_{1D} + \sqrt{2} \lambda_S \mu\right)^2}{m_{SR}^2} - \frac{\left(g_2 m_{2D} - \sqrt{2} \lambda_T  \mu\right)^2}{m_{TR}^2} \nn\\
\delta \lambda_3^{(tree)}  \simeq &  \, \, \, \,     \frac{g_Y^2 m_{1D}^2 - 2\lambda_S^2 \mu^2}{m_{SR}^2}-  \frac{g_2^2 m_{2D}^2 - 2\lambda_T^2 \mu^2}{m_{TR}^2} \nn  \\
\delta \lambda_4 ^{(tree)} \simeq & \, \, \, \,  \frac{2 g_2^2 m_{2D}^2 - 4 \lambda_T^2 \mu^2}{m_{TR}^2}\,, \nn\\
\label{deltaLambdaTree}
\end{align}

The effect on the quartic scalar potential can be written:
 \begin{eqnarray}
\delta V_{4\Phi}^{(tree)}  &=& \delta \lambda_{|0_1,0>}^{(tree)}  |0_1,0\rangle  +   \delta \lambda_{|0_2,0>}^{(tree)}   |0_2,0\rangle  +  \delta \lambda_{|1,0>}^{(tree)}  |1,0\rangle  +   \delta \lambda_{|2,0>}^{(tree)}   |2,0\rangle \, .
\label{deltareparam2HDM} 
\end{eqnarray}
The two singlet coefficients get corrections 
\begin{align}
\delta \lambda_{|0_1,0>}^{(tree)} \simeq & -2 \lambda_S^2  \frac{\mu^2}{m_{SR}^2} - g_2^2 \frac{ m_{2D}^2}{m_{TR}^2}  \\
\delta  \lambda_{|0_2,0> }^{(tree)}\simeq &   \frac{1}{\sqrt{3}} \left[ g_Y^2 \frac{ m_{1D}^2 }{m_{SR}^2} - 2 g_2^2 \frac{ m_{2D}^2 }{m_{TR}^2}  + 6 \lambda_T^2 \frac{\mu^2}{m_{TR}^2} \right]
\label{deltaLambdaTreeSU2R}
\end{align}
and, by themselves do not contribute to misalignment. The breaking of the $SU(2)_R$ symmetry shows up through the appearance of new terms in the scalar potential:
\begin{align}
\delta \lambda_{|1,0>}^{(tree)} \simeq & 2 g_2 \lambda_T  \frac{m_{2D} \mu}{m_{TR}^2}  -2 g_Y \lambda_S  \frac{m_{1D} \mu}{m_{SR}^2}  \nn \\
\simeq & \sqrt{2} g_2^2  \frac{m_{2D} \mu}{m_{TR}^2}  -\sqrt{2} g_Y^2   \frac{m_{1D} \mu}{m_{SR}^2}  \nn \\
\delta  \lambda_{|2,0> }^{(tree)}\simeq &  \sqrt{ \frac{2}{3}} \left[g_Y^2 \frac{ m_{1D}^2 }{m_{SR}^2} + g_2^2 \frac{ m_{2D}^2 }{m_{TR}^2} \right]
\label{deltaLambdas}
\end{align}
that preserve the subgroup $U(1)'_R$ as expected as the scalar potential results from  integrating out the adjoints which have zero $U(1)'_R$ charge. For a numerical estimate, we take the example of values used in \cite{Benakli:2018vqz} with $m_{SR} \simeq m_{TR} \simeq 5$  TeV,   $m_{1D}\simeq m_{1D} \simeq \mu \simeq 500$ GeV,  $g_Y\simeq 0.37$ and $g_2 \simeq 0.64$.  This gives 
\begin{align}
\delta \lambda_{|1,0>}^{(tree)} \simeq & 4 \times 10^{-3},  \qquad
\delta  \lambda_{|2,0> }^{(tree)}\simeq 4.5 \times 10^{-3}&  
\label{deltaLambdaTreeSU2R2}
\end{align}
which show that the contribution to $Z_6$ is very small and they will not be discussed further below.

Next, we consider the misalignment from quantum corrections. Radiative corrections to different couplings are generated when supersymmetry is broken inducing mass splitting between scalars and fermionic partners. This happens for instance through loops of the adjoint scalar fields $S$ and $T^a$. However, these scalars are singlets under the $SU(2)_R$ symmetry and at leading order, when their couplings $\lambda_S$ and $\lambda_T$ are given by their $N=2$ values, their interactions with the two Higgs doublets preserve $SU(2)_R$.  Therefore we do not expect them to lead to any contribution to $Z_6$. In fact, explicit calculations of these loop diagrams was performed in eq. (3.5) of \cite{Benakli:2018vqz} and it was found that when summed up their contribution to $Z_6$ cancels out. This unexpected result is now easily understood as the consequence of the $SU(2)_R$ symmetry.

Let's denote the auxiliary fields  $D^a$ for the gauge fields $A^a$ and $F_\Sigma^a$ for the corresponding adjoint scalars $\Sigma^a \in \{ S,T^a \}$ of $U(1)_Y$ and $SU(2)$. Then the set:
\begin{align}
( F_\Sigma^a \quad ,  \qquad
D^a \quad , \qquad  {F_\Sigma^a}^*  )
\label{Auxil}
\end{align}
forms a triplet of $SU(2)_R$. This enforces the equalities $\lambda_S=  g_Y/{\sqrt{2}}$ and $ \lambda_T = g_2/{\sqrt{2}}$ in eq. (\ref{LSTN2}) whose violation by quantum effects translates into breaking of $SU(2)_R$. First, consider the correction due to running of the above couplings. This accounts for the violation of  $N=2$ induced relations (\ref{LSTN2}) due to radiative corrections from $N=1$ chiral matter. As $\lambda_1$ and $\lambda_2$ are affected in the same way, we have $\delta \lambda_{|0_1,0>}^{(2\rightarrow 1)} =0$, and using (\ref{lambdaofSU(2)}), we get:
 \begin{eqnarray}
\delta  Z_6^{(2\rightarrow 1)} &=&  - \frac{\sqrt{6}}{2}  \, \,  s_{2\beta}  \, \,  c_{2\beta}   \, \,  \delta   \lambda_{|2,0>}^{(2\rightarrow 1)} \nn \\
&= &  -  \frac{1}{2}  \, \,\frac{ t_\beta ( t_\beta^2- 1)}{( t_\beta^2 +1)^2}  \, \,\left[ (2 \lambda_S^2- g_Y^2  )  + ( 2 \lambda_T^2 -g_2^2 ) \right] 
\label{Z6N2-1}
\end{eqnarray}

Another source of misalignment comes from the  $N=1 \rightarrow N=0$ mass splitting in chiral superfields. The $SU(2)_R$ symmetry is broken by the different Yukawa couplings to the two Higgs doublets, which for $t_\beta \sim \mathcal{O}(1)$ will be dominated by the top Yukawa.  The stops contributes mainly by correcting $\lambda_2$.
\begin{align}
 \delta  \lambda_2 \sim &  \frac{3y_t^4}{8\pi^2} \log \frac{m_{\tilde{t}}^2}{Q^2} 
\label{toplambda2}
\end{align}
where $y_t$, $m_{\tilde{t}}$ are the top Yukawa and stop masses, respectively, while $Q$ is the renormalisation scale. This leads to a $\delta  Z_6^{(1\rightarrow 0)}$ induced by $ \delta \lambda_{|1,0>}^{(stops)} \sim \sqrt{3} \delta   \lambda_{|2,0>}^{(stops)}$ in (\ref{mis-Z6-tree-SU(2)}). Thus, chiral matter  through their Yukawa couplings contribute  to $Z_6$ with both of $|1,0\rangle$ and $|2,0\rangle$ combinations of doublets. The contributions $\delta  Z_6^{(1\rightarrow 0)}$ and $\delta  Z_6^{(2\rightarrow 1)}$ have similar strength but opposite sign so that they lead to a small misalignment compatible with LHC bounds as was shown in \cite{Benakli:2018vqz} by explicit computation.

Note that the the origin of misalignment is not the quadratic terms breaking $SU(2)_R$. In section 3, we explained how this breaking of $SU(2)_R$ does not imply loss of alignment but fixes $\tan \beta$. The misalignment comes the presence of chiral matter and the most important single contribution arises from a large top Yukawa coupling.

\section{Conclusions}

In the 2HDM, the LHC experiments data require one the Higgs squared-mass matrix eigenstates to be aligned with the SM-like direction. If this alignment comes without decoupling, then the additional scalars in the electroweak sector are subject to milder constraints. They might have masses in an energy range that can be reached in future LHC searches, leaving open the possibility of new discovery of fundamental scalars. This alignment was not only achieved but also predicted at tree-level in \cite{Antoniadis:2006uj}. This success calls for understanding the main mechanism behind it. We have shown in this work that it is an $SU(2)_R$ R-symmetry that acts as a Higgs family symmetry in the quartic scalar potential and enforces an automatic alignment. Another new result of this work is that we have written the CP-even Higgs squared-mass matrix off-diagonal  element as a linear combination of the coefficients of non-singlet of $SU(2)_R$ representations. These are generated in the quartic potential by tree level threshold and loop corrections. Their numerical values have been computed in \cite{Benakli:2018vqz} where it was proven that the alignment is preserved to an unexpected precision level.  A new way of expressing the observables is presented here.  We have found that using $SU(2)_R$ symmetry allows to shed light on some of the existing results. For instance, we understand the origin of the cancellation between loop contributions of the adjoint scalars. When interested by alignment, only some contributions need to be evaluated or computed explicitly in the future: those contributing to coefficients of particular combinations of terms in the potential that break the $SU(2)_R$ symmetry. This sheds light not only on how the alignment is realized here, but also on why $N=2$ realizations of other models as the MRSSM attempted  in \cite{Benakli:2018vqz} have not been successful. The extension of the quartic potential to an $N=2$ sector with a particular attention to preserving the $SU(2)_R$ symmetry at tree-level can be a way to pursue in order to implement alignment in extended Higgs sector models.

%\vskip.1in
\noindent
\section*{Acknowledgments}
 We are grateful to  P.~Slavich for discussions. We acknowledge the support of  the Agence Nationale de Recherche under grant ANR-15-CE31-0002 ``HiggsAutomator''. This work is also supported by the Labex ``Institut Lagrange de Paris'' (ANR-11-IDEX-0004-02,  ANR-10-LABX-63)).

%\newpage

\noindent

\providecommand{\href}[2]{#2}\begingroup\raggedright\endgroup

\end{document}